\documentstyle[preprint,prl,aps,epsfig]{revtex}
\begin{document}
%srklein@lbl.gov, rlvogt@lbl.gov

\title{Deuteron Photodissociation in Ultraperipheral \\ 
Relativistic Heavy-Ion on Deuteron Collisions}

\author{Spencer Klein$^a$ and Ramona Vogt$^{a,b}$} 
\address{$^a$Lawrence Berkeley National Laboratory, Berkeley, CA, 94720\break
$^b$University of California, Davis, CA, 95616}

\break 
\maketitle
\vskip -.2 in
\begin{abstract}
\vskip -.2 in 

In ultraperipheral relativistic deuteron on heavy-ion collisions, a
photon emitted from the heavy nucleus may dissociate the deuterium
ion.  We find deuterium breakup cross sections of 1.38 barns for
deuterium-gold collisions at a center of mass energy of 200 GeV per
nucleon, as studied at the Relativistic Heavy Ion Collider, and 2.49
barns for deuterium-lead collisions at a center of mass energy of 6.2
TeV, as proposed for the Large Hadron Collider.  This cross section includes an
energy-independent 140 mb contribution from hadronic diffractive
dissociation.  At the LHC, the cross section is as large as that of
hadronic interactions. The estimated error is 5\%.  Deuteron
dissociation could be used as a luminosity monitor and a `tag' for
moderate impact parameter collisions.

\end{abstract}
\narrowtext

Deuterium-heavy ion collisions are of considerable interest at heavy
ion colliders.  Technically, they are much easier than proton-ion
collisions because the deuteron charge-to-mass ratio, $Z/A$, is
similar to that of heavy ions, greatly simplifying the magnetic optics
around the collision point.  Further, at the LHC, the matching $Z/A$
of the two beams means that the nucleon-nucleon center-of-mass frame
is closer to the lab frame than it would be in $pA$ collisions.  The
similarity in proton:neutron ratios will also simplify the comparison
of data from $dA$ and $AA$ collisions. For these reasons, $dA$
collisions may be preferred over $pA$.

Ultraperipheral $dA$ collisions are also of interest.  The strong
electromagnetic field of the heavy ion acts as an intense photon beam
which strikes the deuterium, producing very high energy $\gamma d$
collisions.  In $AA$ collisions, there is a two-fold ambiguity over
which ion emitted the photon.  However, in $dA$ collisions, the photon
almost always comes from the heavy ion, allowing a clean determination
of the photon energy based on the final state rapidity.

Some ultra-peripheral reactions are unique to $dA$.  Here, we consider
one example, photodissociation of deuterium.  This reaction has a very
large cross section and can serve as a luminosity monitor and as a
`tag' for moderate impact parameter UPCs.  The photodissociation cross
section is
\begin{equation}
\sigma_{\rm diss} = \int dk \, {dN\over dk} \, \sigma_d(k)
\label{crosssec}
\end{equation}
where $dN(k)/dk$ is the photon flux from the heavy ion and
$\sigma_d(k)$ is the photon-deuteron breakup cross section.

The photon flux emitted by the heavy ion is obtained from the
Weizsacker-Williams approach.  The flux, integrated over impact
parameters, $b$, greater than $R_{\rm min}$ is \cite{reviews}
\begin{equation}
{dN\over dk}  = {2Z^2\alpha \over \pi k} \left[ xK_0(x)K_1(x) - {x^2\over 2} 
\big(K_1^2(x) - K_0^2(x)\big) \right]
\end{equation}
where $\alpha$ is the fine structure constant, $K_0$ and $K_1$ are
Bessel functions and $x=k R_{\rm min}/\hbar c \Gamma$.  Here
$\Gamma=2\gamma^2-1$ where $\gamma$ is the Lorentz boost of a single
beam.  The minimum radius, $R_{\rm min} = R_A + R_d$, is required to
eliminate collisions that include hadronic interactions.  We take 2.1
fm for the deuteron radius \cite{drad} and assume $R_A = 1.2A^{1/3}$ for the
heavy ion.  Thus $R_{\rm min} = 9.08$ fm for $d$Au and 9.21 fm for $d$Pb
collisions.  We will discuss the sensitivity to $R_{\rm min}$ later.

\begin{figure}
\setlength{\epsfxsize=0.95\textwidth}
\setlength{\epsfysize=0.55\textheight}
\centerline{\epsffile{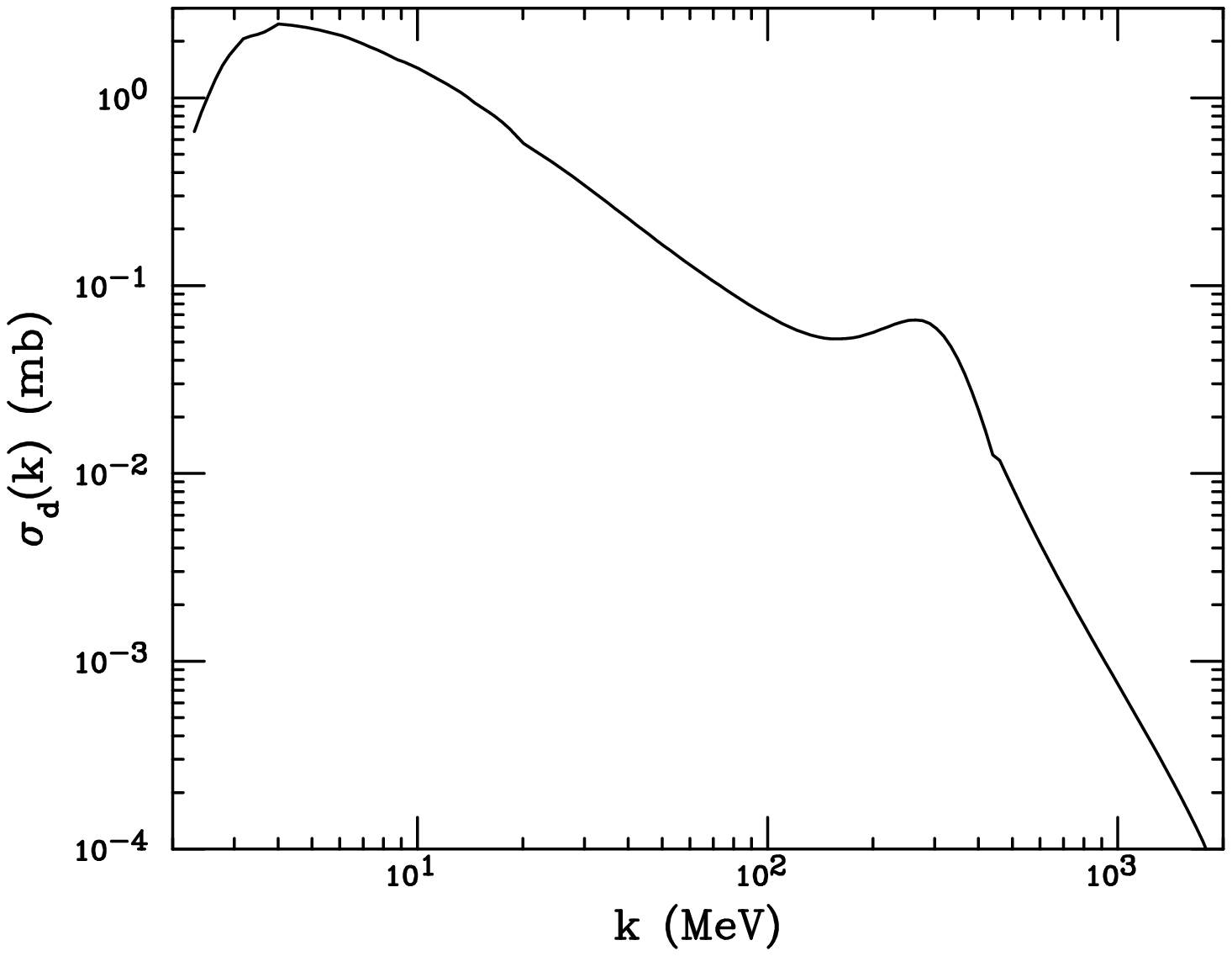}}
\caption[]{Cross section for deuteron photodissociation
as a function of photon energy, in the deuteron rest frame.}
\label{fig1}
\end{figure}

Since the photon spectrum scales as $1/k$, deuteron breakup is
dominated by interactions near threshold, 2.23 MeV in the deuteron
rest frame.  There have been a number of measurements of deuteron
breakup by low energy photons.  Our breakup cross section,
Fig.~\ref{fig1}, is based on measurements at 2.754 MeV \cite{d1}, 4.45
MeV \cite{d2}, $5.97<k<11.39$ MeV \cite{d3}, and $15<k<75$ MeV
\cite{d4}.  For $k<$ 4 MeV, we use the cross sections calculated with
``Approximation III'' in Table 1 of Ref.~\cite{d6} since this
approximation matches the 2.754 MeV data point. In the region
$20<k<440$ MeV, we rely on a fit to the data \cite{d5}.  For $440<k <
625$ MeV, we use a slightly earlier fit \cite{d7}.  
We extrapolate this fit to 2 GeV with some loss in accuracy. At
higher energies, QCD counting rules predict that $\sigma_d$
should drop rapidly with energy \cite{brodsky}.  Thus we neglect
energies above 2 GeV.

\begin{figure}
\setlength{\epsfxsize=0.95\textwidth}
\setlength{\epsfysize=0.55\textheight}
\centerline{\epsffile{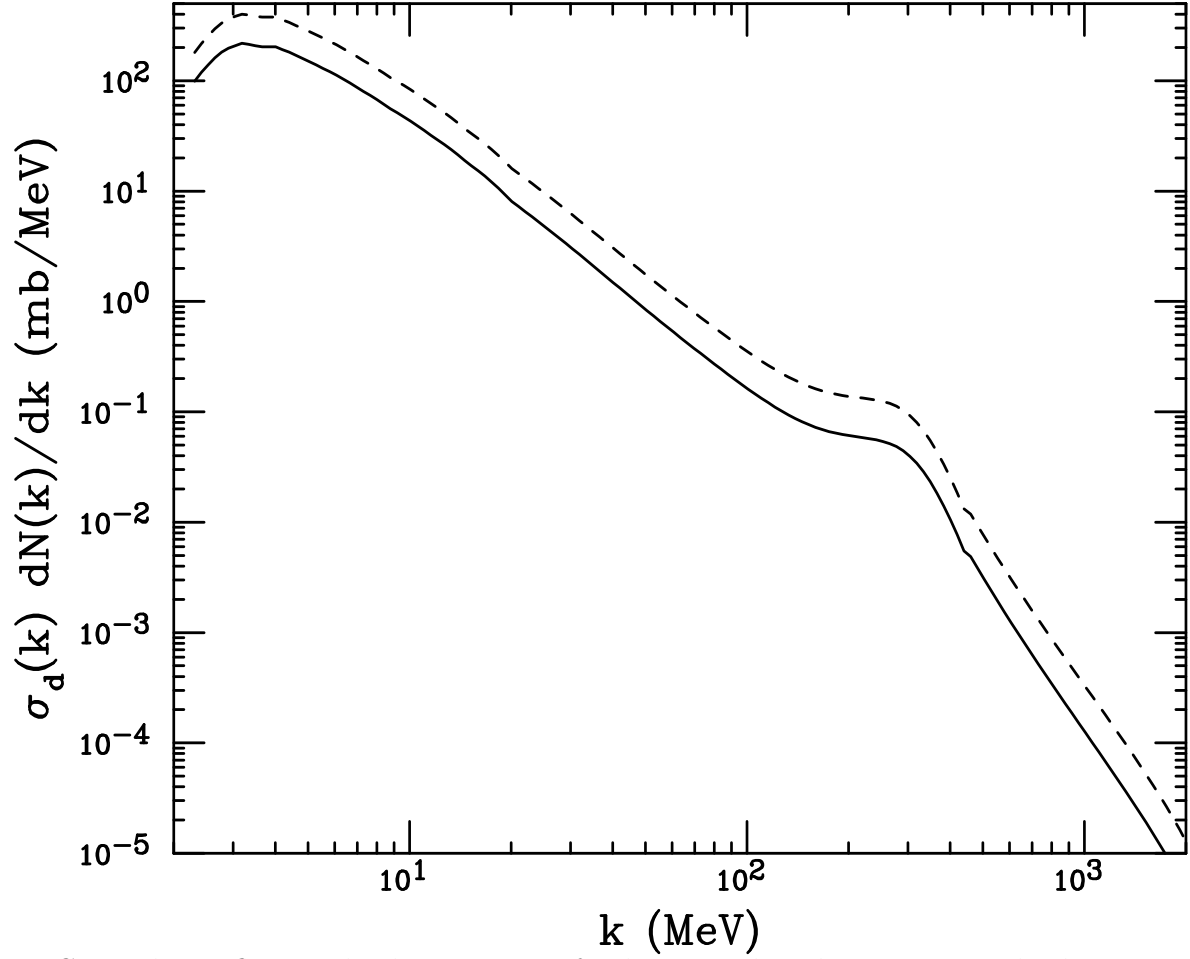}}
\caption[]{Photon-flux weighted cross section for deuteron
photodissociation, in the deuteron rest frame, for $d$Au at
RHIC (solid line) and $d$Pb at the LHC (dotted line).}
\label{fig2}
\end{figure}

The integrand of Eq.~(\ref{crosssec}) is shown in Fig.~\ref{fig2}.
The integrated photodissociation cross section is $1.24$ barns for
deuterium-gold collisions at 200 GeV/nucleon center of mass energy, as
currently studied at the Relativistic Heavy Ion Collider (RHIC).  At a
luminosity of $4\times 10^{28}$ cm$^{-2}$s$^{-1}$\cite{RHIC}, this is
50,000 interactions/sec.  For deuterium-lead collisions at 6.2
TeV/nucleon, as may be studied at the Large Hadron Collider (LHC), the
calculated cross section is $2.35$ barns.  At a luminosity of $2.7
\times 10^{29}$ cm$^{-2}$s$^{-1}$ \cite{dpblum}, there will be 650,000
interactions/sec.  These cross sections are comparable to the
calculated hadronic cross sections at RHIC, $2.26\pm 0.10$ barns
\cite{dima} and $2.37^{+0.13}_{-0.12}$ barns\cite{miller}, and
slightly larger than the hadronic cross sections at the LHC. The
quoted 4\% uncertainty in Ref.~\cite{dima} includes only the deuteron
wave function.

The accuracy of the photodissociation cross section depends on the
data from which it is derived.  The most important photon energy range
is below $10$ MeV.  Above 5.9 MeV, the data are quite accurate with
uncertainties well below 5\%.  The 2.754 MeV data point also has only
a 3.2\% error.  Unfortunately, at intermediate energies, the only data
are from a 1952 measurement at 4.45 MeV with a 7\% uncertainty.  The
theoretical extrapolation from lower energies should be more accurate
than the data in this range.  We estimate that the uncertainty in
$\sigma_d$ contributes a 4\% error to the total $dA$ cross sections.

Another uncertainty arises from the truncation of the photon spectrum.
With the rapidly falling spectrum, stopping the calculation at 500 MeV
reduces the photodissociation cross section by less than 0.1\%.  We
estimate that the truncation introduces an error of less than 0.5\%
into the calculation.

The other important uncertainty in the cross section is in $R_{\rm
min}$.  The charge radii of heavy ions are well measured \cite{Vvv}
but the radii of the matter distribution may be $0.1-0.3$ fm
larger\cite{pbars}.  In addition, heavy ions can have non-negligible
densities even at quite large distances.  The deuteron wave function is
complex, making an accurate geometric calculation quite difficult.
However, because the photon energies are so low, the choice of $R_{\rm
min}$ is not important.  For gold at RHIC, increasing $R_{\rm min}$ by
2 fm only reduces the cross section by 1.8\% while for lead at the
LHC, a 2 fm increase reduces it by 1.1\%.

In addition to photodissociation, diffraction can induce
deuteron dissociation.  The heavy nucleus absorbs part
of the deuteron wave function, leaving two independent nucleons in the 
outgoing state.  For a completely black (absorptive) target nucleus with
radius $R_A$, the diffractive dissociation
cross section is \cite{glauber,akhieser} 
\begin{equation}
\sigma_{\rm diff} = {\pi\over 3}\big( 2 \ln{(2)} - 0.5 \big) R_A R_d \, \, .
\end{equation}
We find $\sigma_{\rm diff} = 136$ mb for gold and $\sigma_{\rm diff} =
139$ mb for lead.  Since
heavy nuclei are not completely black, this approach probably
overestimates the cross sections slightly.  However, since diffractive
breakup is a small fraction of the total cross section, $\sigma = \sigma_{\rm
diss} + \sigma_{\rm diff}$, we do not
correct for the partial transparency.
 
We estimate the overall uncertainty in the dissociation cross section
to be less than 5\%, comparable to that of the hadronic deuterium-ion
cross sections.  The uncertainties in the hadronic radii of heavy ions
and the reaction geometry are at least as problematic as for hadronic
interactions.

Experimentally, photodissociation has a clean signature: a proton and
a neutron with roughly the beam momenta and no other visible reaction
products.  Other photonuclear interactions can break up the deuteron
and create additional particles but they represent a small fraction of
the photodissociation cross section.  The resulting neutron and proton
can be detected in a zero degree calorimeter \cite{zdc} and a forward
proton calorimeter respectively.  Because of the small excitation
energies, even small calorimeters will have good acceptance for the
reaction products.

One final experimental issue is background.  Deuteron-beam gas
interactions might mimic photodissociation.  However, in beam-gas
interactions the proton or neutron will usually lose a large fraction
of its energy, allowing these events to be rejected.  The beam-gas
background can be measured by momentarily separating the beams to stop
the collisions.  With this check, deuteron breakup would be a useful
calibration reaction for van der Meer scans of absolute luminosity
\cite{vdm}, and, in routine operations, as a luminosity monitor.
Because of the high rates, a neutron calorimeter alone will likely
suffice for dissociation studies.  In many respects, the use of
deuteron photodissociation parallels the use of mutual Coulomb
dissociation in heavy ion collisions\cite{breakup}.

\begin{figure}
\setlength{\epsfxsize=0.95\textwidth}
\setlength{\epsfysize=0.3\textheight}
\centerline{\epsffile{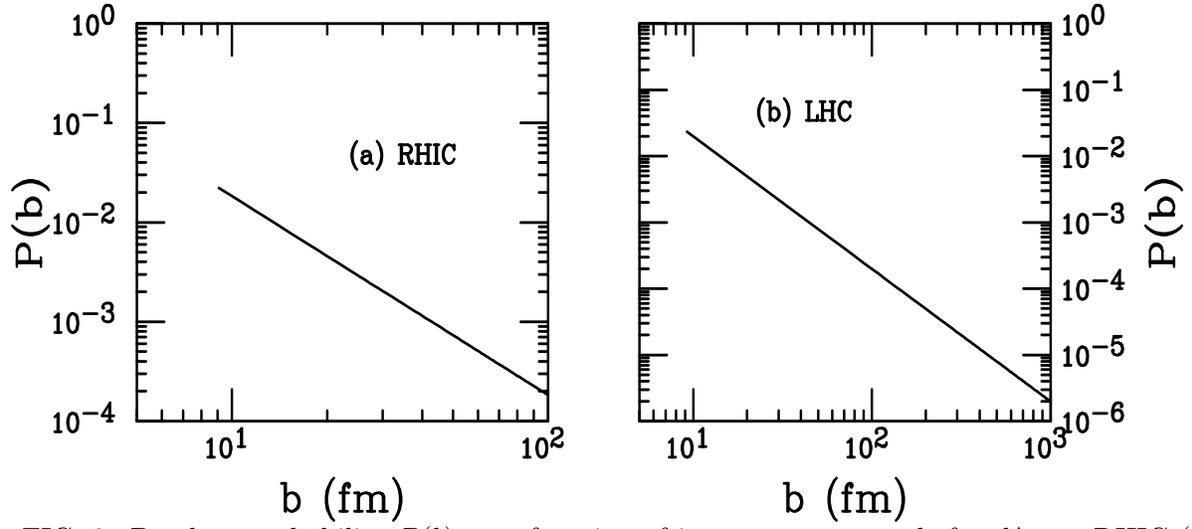}}
\caption[]{Breakup probability $P(b)$ as a function of impact
parameter $b$, for $d$Au at RHIC (a) and $d$Pb at the
LHC (b).}
\label{fig3}
\end{figure}

Photodissociation can be used as an impact parameter tag for studying
other ultraperipheral collisions, as is done for mutual Coulomb
dissociation \cite{tags}.  The final state neutron (or proton)
provides a tag of a moderate impact parameter encounter.  The
probability of photodissociation at impact parameter $b$, $P(b)$, is
calculated using the impact-parameter dependent photon
flux \cite{vectors}:
\begin{equation}
P(b) = \int  {d^3 N\over dkd^2b} \sigma_d(k) dk,
\end{equation}
where
\begin{equation}
{d^3N\over dkd^2b} = {Z^2\alpha k\over\Gamma^2\pi^2}
\big(K_1^2(x) +K_0^2(x)/\Gamma^2\big)
\end{equation}
and here $x=kb/\hbar c\Gamma$.  As long as $x < 1$ for the important
photon energies, $2 < k < 10$ MeV, $P(b) \approx 1/b^2$ and
$d\sigma/db\approx 1/b$.  
This condition holds for $b< 0.4$ nm at RHIC and $b<0.4$ $\mu$m at the LHC.
Photodissociation can thus occur at extremely large impact parameters!
The breakup probability as a function of $b$ is given in
Fig.~\ref{fig3}.  At an impact parameter of 10 fm, the probability of
deuteron breakup is 1.8\% for $d$Au collisions at RHIC, dropping to
0.1\% at 45 fm.  These probabilities are somewhat lower than for
mutual Coulomb breakup in $AA$ collisions but should still be a useful
tag.

Diffractive dissociation can occur when there is some overlap between the
ion and deuteron wave functions.  Including this contribution would increase
the total breakup probabilities slightly for $b < 20$ fm.

In conclusion, we find that the cross section for deuteron breakup in
$d$Au collisions at RHIC is 1.38 barns, while at the LHC, the cross
section for dissociation in $d$Pb collisions is 2.49 barns.  Both
cross sections have an estimated error of 5\%.  This reaction has a
well determined cross section and a clean signature, giving it utility
as a `calibration' for luminosity measurement and monitoring and as a
trigger for other ultraperipheral collisions.

We thank Kai Hencken and Gerhard Baur for useful discussions about
diffractive dissociation.  This work was supported by the US DOE,
under contract DE-AC-03-76SF00098.

\newpage

\end{document}